# Active Control of Mode Crossover and Mode Hopping of Spin Waves in a Ferromagnetic Antidot Lattice

Samiran Choudhury,[†] Sudip Majumder,[†] Saswati Barman,[†,§] YoshiChika Otani,[‡,§§] and Anjan Barman[*,†]

[†]Department of Condensed Matter Physics and Material Sciences, S. N. Bose National Centre for Basic Sciences, Block JD, Sector III, Salt Lake, Kolkata 700 106, India

[§]Institute of Engineering and Management, Sector V, Salt Lake, Kolkata 700 091, India

[‡]Institute for Solid State Physics, University of Tokyo, 5-1-5 Kashiwanoha, Kashiwa, Chiba 277-8581, Japan

[§§]RIKEN-CEMS, 2-1 Hirosawa, Wako, Saitama 351-0198, Japan

[*]E-mail: abarman@bose.res.in

**ABSTRACT**

Active control of spin-wave dynamics is demonstrated using broadband ferromagnetic resonance in two-dimensional $Ni_{80}Fe_{20}$ antidot lattices arranged in hexagonal lattice with fixed lattice constant but varying antidot diameter. A strong modification in the spin-wave spectra is obtained with the variation in the antidot diameter as well as with the strength and orientation of the bias magnetic field. A broad band of modes is observed for the lattice with higher antidot diameter which decreases systematically as the antidot diameter is reduced. A crossover between the higher frequency branches is achieved in lattices with higher antidot diameter. In addition, the spin-wave modes in all lattices show a strong six-fold anisotropic behaviour due to the variation of internal field distribution as a function of the bias-field orientation. A mode hopping-like behavior is observed in the angular dispersions of spin-wave spectra for samples having intermediate hole diameters. Micromagnetic simulations qualitatively reproduce the experimentally observed spin-wave modes and the simulated mode profiles reveal the presence of extended and quantized standing spin-wave modes in these lattices. These observations are significant for large tunability and anisotropic propagation of spin waves in GHz frequency magnetic devices.

## I. INTRODUCTION

Ferromagnetic (FM) antidot (hole) lattices (ADLs)[1], *i.e.* periodically arranged holes embedded in a ferromagnetic thin film are artificial crystals, which are fabricated by structuring known ferromagnetic materials at different length scales. Exploitation of their dynamic responses over a broad temporal and spatial regimes can offer various exciting properties. These magnetic nanostructures form the basis of future technologies including magneto-photonic crystals[2,3] and ultrahigh density storage device. They also possess exciting prospects in the field of magnonics as magnonic crystals (MCs)[4] where spin waves (SWs) are used to carry and process the information in the microwave band analogous to photonic and phononic crystals. They can play a key role in building nanoscale magnonic devices for GHz frequency communication[5], waveguides[6], phase shifters[7], filters[8], interferometers[9], spin-wave logic devices[10] and spin-wave nano-optics[11] with spin waves in multiple connected magnon waveguides. More recently, the filled antidot lattices also have gained a great interest as bi-component MCs[5,12] (BMCs) for the additional tunability in the magnonic band structures due to the differences between the magnetic parameters of two magnetic materials in those systems. The upsurge in the nanofabrication and detection techniques with improved spatio-temporal resolution boost in the study of high-frequency magnetization dynamics in a variety of such nanoscale antidot lattices. One important problem of magnonics research is to tune the magnonic spectra and the band structures of the MCs by varying its physical parameters such as antidot shape[13], lattice symmetry[14,15], lattice constant[16], base material[17] (2014)] as well as strength and orientation of the bias magnetic field[18,19] which can greatly affect the SW dynamics.

In the past few years, the high-frequency magnetization dynamics of magnetic ADLs have been explored by various experimental and numerical methods[18-23]. Initial studies on ADLs showed attenuation of uniform ferromagnetic resonance mode due to the excitation of non-uniform in-plane SW mode[24] and also the pattern induced splitting of surface and volume modes[25] were observed. Later, field dependent localization of SW mode, SW confinement and field-controlled propagation of SWs[26,27] as well as the formation of magnonic miniband with large SW velocities[28,29] have been observed. Recently, the dispersive and entangled SWs between the antidots[30] and anisotropic propagation and damping of SWs[18] were also observed due to the magnetic field-induced SW guiding in a network of interconnected nanowires. Further works showed high-symmetric magnonic modes having a linear bias magnetic field dependence for perpendicularly magnetized ADLs[31], and conversion of quantized SWs to propagating ones by varying the bias magnetic field orientation[19]. Micromagnetic simulations reported[32] the effect of the antidot shape on the magnonic band structure in exchange-dominated one-dimensional magnonic waveguides. The effects of lattice defects[33] and broken translational symmetry[34] have also attracted great interest due to their unique properties. However, there are very few reports exploring the effect of the size of the antidots[35], which can play a key role in determining the nature of SW propagation and confinement due to the modulation of the magnonic band gaps in such MCs.

Here, we investigate the SW dynamics of 2-D arrays of circular shaped $Ni_{80}Fe_{20}$ antidots arranged in hexagonal lattice with fixed lattice constant but varying antidot diameter by manipulating the strength and orientation of the in-plane bias magnetic field. A drastic variation in the magnonic spectra is observed when the diameter of the antidot is systematically tuned. A rich band of SW modes for highest antidot diameter reduced systematically to a fewer number of modes with the reduction in antidot diameter. Further, a mode crossover with bias-field strength and mode hopping-like behaviour with bias-field orientation are observed, which are explained with the aid of micromagnetic simulations.

**II. SAMPLE FABRICATION AND EXPERIMENTAL TECHNIQUE:**

Circular shaped antidots are patterned in a 20-nm-thick $Ni_{80}Fe_{20}$ (NiFe) film by using a combination of e-beam lithography (EBL), e-beam evaporation (EBE) and ion milling[15]. The antidots with variable diameters (*d*) of 140 (D1), 240 (D2), 340 (D3), 440 (D4) nm and fixed lattice spacing (*a*) of 700 nm are arranged in hexagonal lattice symmetry with total array dimension of 25 µm × 250 µm, as shown in the scanning electron micrographs of Fig. 1(a). A co-planar waveguide (CPW) made of Au with a thickness of 150 nm was deposited on top of the array for the broadband ferromagnetic resonance measurement. The width and length of the signal line of this CPW were 30 µm and 300 µm, respectively so that the whole array could remain under the centre of the signal line of CPW which has a nominal characteristic impedance of 50 Ω.

The measurement of SW spectra from the samples was performed using a broadband ferromagnetic resonance (FMR) spectrometer[36] consisting of a vector network analyzer (Agilent PNA-L, Model No.: N5230C, frequency range: 10 MHz to 50 GHz) and a high-frequency probe station along with a nonmagnetic G-S-G type pico-probe (GGB Industries, Model No. 40A-GSG-150-EDP). The system has an inbuilt electromagnet within the probe station generating a bias magnetic field ($H_{ext}$) of ± 1800 Oe. The electromagnet is mounted on a high-precision rotary mount which enables the electromagnet to rotate over 360° angle within the plane of the sample. The sample is viewed with the help of a microscope and illumination set-up. Microwave signal with variable frequency is launched into the CPW using the pico-probe through a high frequency and low noise coaxial cable (Model No.: N1501A-203). The CPW is shorted at one end and the back-reflected signal is collected by the same probe to the analyzer. Absorption of the ongoing and returning signals at various SW frequencies produces the characteristic SW spectrum of the sample. The real and imaginary parts of the scattering parameter in reflection geometry measured at various magnetic fields are subtracted from its value at the maximum bias magnetic field (reference spectrum), and hence, the SW spectra are obtained (see Supplemental Material for further details about sample fabrication and experimental method).

**III. EXPERIMENTAL RESULTS AND MICROMAGNETIC SIMULATION**

The real parts of the forward scattering parameter, *i.e.* $S_{11}$ for the samples with varying antidot diameter (*d*) are shown in Fig. 1(b) at a bias magnetic field $H_{ext}$ = 800 Oe applied at an azimuthal angle $\phi$ = 0°, while their bias field dependent SW dispersion spectra measured at $\phi$ = 0° are represented as surface plots in Fig. 2. The SW dynamics get drastically modified with the variation of *d*. For D1, two distinct SW modes are obtained, while, in case of D2, the number of modes is increased to three. On the contrary, rich multimode spectra are observed in D3 and D4 consisting of total six and eight modes. Interestingly, in D4, with the decreasing bias field, we observe a significant crossover (marked by the red dotted box in the field dispersion of D4 in Fig. 2) between the two higher frequency branches (closed and open diamond marked modes 7 and 8, respectively) at an intermediate field value of $H_{ext}$ ~ 525 Oe. A crossover is also observed in D3 as shown by the black dotted box in Fig. 2 between modes 5 and 6 (closed and open diamond marked modes) at a much lower field $H_{ext}$ ~ 170 Oe, while this phenomenon is completely absent in both D1 and D2. Also, the lowest frequency mode M1 in both D3 and D4 vanishes above a certain bias field value (~ 700 Oe for D3 and ~ 430 Oe for D4). The bias field dispersion curves of D3 and D4 also reveal that there are some SW modes *e.g.*, M2 in D3 and M5 in D4 which are present at lower bias field. This is due to the deviation from the uniform magnetization state due to the increase in the overlapping between the demagnetizing regions around the antidots.

We have investigated the origin of the SW modes by performing micromagnetic simulations using the OOMMF software[37] for these samples consisting of 2 × 2 arrays of hexagonal unit cells for each sample where the two-dimensional periodic boundary condition has been incorporated to consider the large areas of the experimentally studied arrays. We have discretized each sample into rectangular prism-like cells with dimensions $4 \times 4 \times 20$ nm$^3$. The value of exchange stiffness constant and saturation magnetization used in the simulation for NiFe are $A_{NiFe} = 1.3 \times 10^{-6}$ erg/cm[38], $M_{NiFe} = 850$ emu/cc, while the damping coefficient $\alpha_{NiFe} = 0.008$ is used for NiFe during the dynamic simulations. The value of gyromagnetic ratio $\gamma = 18.1$ MHz/Oe and the magnetocrystalline anisotropy $K = 0$ are considered for NiFe. Here, the material parameters, *i.e.* $M_s$, $\gamma$ and $K$ for NiFe were extracted from the Kittel fit of the bias-field dependent SW absorption spectra of NiFe thin film (see Supplemental Material, Fig. S1). The dynamic simulations were carried out by first performing a static magnetic configuration under a bias magnetic field in the experimental geometry and then by applying a pulsed magnetic field. The details of the static and the dynamic simulations are described elsewhere[39].

Figure 1(c) shows that the experimental data were reproduced qualitatively well using the micromagnetic simulations which were plotted at $H_{ext} = 800$ Oe, and also in Fig. 2 as represented by filled symbols for all the lattices. The slight quantitative disagreements between experimental and simulated results can occur due to the deviation of the simulated samples and conditions from the experimental ones. The general deviation in the dimensions as observed in the experimental samples has already been incorporated in the simulated samples, although the precise edge deformations are not possible to include in the finite difference method based micromagnetic simulations.

In Fig. 3(a), we have shown the simulated static magnetic configurations for D1 and D2 at $H_{ext} = 800$ Oe ($\phi = 0°$), while, in order to underpin the origin of the mode crossover observed in both D3 and D4, we have presented their static magnetization states at two different bias field values, $H_{ext} = 200$ and 800 Oe. The static magnetization maps of D3 and D4 at $H_{ext} = 200$ Oe revealed a strong overlap of the demagnetized regions present between the two antidots situated diagonally (as shown by green dotted boxes) of the lattices, while this overlap reduces systematically with the increase in bias field and vanishes for $H_{ext}$ > 300 Oe in D3 and $H_{ext}$ > 600 Oe in D4, respectively. For a better understanding of the nature of various observed SW modes, we have calculated the spatial distributions of these SW modes by using a homebuilt code[40]. Figure 3(b) shows the phase profiles (corresponding power profiles are shown in the Supplemental Material, Fig. S2) of the SW modes for all samples with varying $d$, calculated at $H_{ext}$ = 800 Oe for $\phi$ = 0°. We have observed different types of extended and quantized standing SW modes due to the formation of confining potentials by the demagnetizing fields around the antidots. So, in order to keep uniformity in describing the nature of the observed standing SW modes, we have assigned quantization number $n$ for the modes forming standing waves in the rhombic-region between the two (horizontally situated) consecutive antidots along x-direction as shown by blue dotted box in the phase profile of M1 for D1. It is clear from Fig. 3(b) that the lowest frequency mode M1 for D1 is extended through the diagonally situated antidots along y-direction. However, considering the rhombic region formed by four antidots, it also forms standing wave mode along x-direction with quantization $n$ = 3, while the other mode M2 has quantized character along x-direction with $n$ = 5. Similarly, in D2, both the lower frequency modes M1 and M2 represent extended nature in the y-direction. Although, M1 and M2 also possess quantized character along x-direction in the rhombic unit with $n$ = 3 but they are in opposite phase with each other. The highest frequency mode M3 of D2 also forms standing wave pattern with $n$ = 11. The mode profiles get substantially modified with further increase in $d$, as all the SWs form standing waves in x-direction for D3 and D4. However, unlike for D1 and D2, some of the SW modes show discontinuity for low or high field values in D3 and D4. To explain that, we have shown the phase maps of the SW modes which are present at $H_{ext}$ = 200 and 800 Oe for D3 and D4. For D3 at $H_{ext}$ = 800 Oe, both M1 and M3 are quantized modes with $n$ = 3 which are opposite in phase, while M4 and M6 also possess quantized character with $n = 5$ and 7, respectively. However, at lower bias field ($H_{ext}$ = 200 Oe), M2 shows quantized nature with $n = 1$. On the other hand, due to the increase in the overlapping of demagnetized regions as evident from the static magnetization profile shown in Fig. 3(a) for D3, a diagonal quantization between the next nearest neighbouring antidots is observed for M5 as shown by the black dotted box. We assign $m$ as the diagonal quantization number where $m = 3$ for M5. In D4, at $H_{ext}$ = 800 Oe, all the modes (M1, M2, M3 and M8) represent quantized nature with $n$ = 1, 1, 3 and 5, respectively, where M1 and M2 are in opposite phase. However, at $H_{ext}$ = 200 Oe, due to strong demagnetizing field, a significant modification is observed where all the SW modes M2-M7 represent quantized modes with $n$ = 1, 3, 3, 3, 3 and 1, respectively with their diagonal quantization value $m$ of 1, 1, 3, 3, 5 and 5 respectively. Here, M4 and M5 are in opposite phase with each other.

We have further calculated the magnetostatic field distributions in the ADLs for bias field $H_{ext}$ = 800 Oe applied at $\phi = 0°$ using the LLG software[41], as shown in Fig. 3(c). It can be clearly seen that the magnetostatic field distribution is modified drastically when $d$ is varied. We have further compared the internal field ($B_{in}$) values of these samples by taking a linescan between two consecutive antidots along x-direction as shown by the black dotted lines in D1-D4. It is found that $B_{in}$ decreases significantly with the increase in $d$, due to the systematic increase in the overlapping between the demagnetizing fields confined around the antidots. This variation of $B_{in}$ has been fitted using a simple parabolic equation which reflects that $B_{in}$ varies as $-d^2$ with the following equation:

$$B_{in} = A(-d^2) + C \qquad (1)$$

where, $C$ is the $B_{in}$ value (~ 10.81 kOe) of NiFe thin film having a thickness of 20 nm. To confirm this behaviour, we have calculated the internal field values for additional simulated samples with $d$ = 40 and 540 nm and the $B_{in}$ values of these samples also follow the same fitting equation (1) (see Supplemental Material, Fig. S3).

To investigate the configurational anisotropy in these samples, we have measured the SW spectra of these samples by varying the in-plane orientation, ($\phi$) of the bias magnetic field at a fixed strength. Figure 4 presents surface plots of the angular dispersion of SW frequencies for all the samples (D1-D4) at $H_{ext}$ = 800 Oe. The solid lines represent the theoretical fits using harmonic functions with different rotational symmetries. The anisotropy of the SW modes varied significantly as $d$ is increased. The lowest frequency mode M1 for D1 shows six-fold rotational anisotropy, while M2 possesses a superposition of a strong six- and weak two-fold anisotropy. The higher frequency anisotropic modes, *i.e.* M2 in D2 and M4 in D3 show a combination of six- and two-fold rotational symmetry. On the other hand, M1 and M8 in D4 also possess six-fold rotational symmetry along with a weak two-fold symmetry, although M8 is in opposite phase with M1. Also, a two-fold anisotropy is observed in D4 for the two intermediate SW modes M2 and M3, while they are in opposite phase with each other.

In contrast, the angular dispersions of the SW modes for D2 and D3 reflect very interesting behaviour as there is a stark modulation in the SW intensity of M1 in both D2 and D3 when the bias field orientation is varied. This phenomenon is more prominent in Fig. 5(a) which have been extracted from Fig. 4, for 120° ≤ $\phi$ ≤ 210°, as marked by the green dotted boxes in D2 and D3, respectively. In D2, for 135° ≤ $\phi$ ≤ 165°, apparently a mode hopping occurs from mode M1 to a new mode M*. However, on a closer look, it appears that the mode splits into a higher power lower frequency mode M*, while M1 also remains with a very low power and M2 possesses a systematic dispersion. This multimodal oscillation for the limited angular range is similar to optical parametric generation, which appears and disappears periodically with an angular period of 60°. Similar behaviour is observed for D3, but here, in addition to the two modes (M* and M4), another mode M3 remains very close to M1. The presence of this additional mode M* has been also confirmed from bias field-dependent SW dispersion curves of D2 and D3 measured at $\phi = 150°$ and $165°$, respectively, as demonstrated in Fig. 5(c). This behaviour is reconfirmed by the SW spectra shown in Fig. 1(b) and Fig. 5(b), which shows the appearance and disappearance of the additional modes with M1 at a regular interval of $60°$.

The evolution of SW phase maps (corresponding power maps are shown in the Supplemental Material, Fig. S4) with $\phi$ in Fig. 6(a) shows that mode M1 in D1 remains almost invariant, *i.e.* extended in the direction perpendicular to the bias field along the NiFe channel between the antidots as marked by the black dotted box in M1 of D1. The quantization along the field direction is modified when $\phi$ is changed from $0°$ to $30°$. However, M2 is converted into an extended mode from quantized mode when $\phi$ is rotated from $0°$ to $15°$ or more. On the other hand, both the SW modes (M1 and M2) in D2 experience a significant transformation as $\phi$ changes to $30°$ and are converted to the extended modes inside the NiFe channel but with different quantization number along the field direction. Similarly, in D3 when $\phi$ is increased, the two lower frequency anisotropic modes M1 and M3 undergo conversion from quantized SW modes to extended modes but with different quantization number inside the NiFe channel, whereas M4 remains spatially invariant with $\phi$. The asterisk marked mode M* in D2 shows extended DE-like nature at $\phi = 30°$, while in D3 it represents quantized behaviour at $\phi = 15°$ similar to M1 although they are in opposite phase with each other as evident from Fig. 6(a, b). Interestingly, the lowest frequency mode M1 in D4 becomes extended mode at $\phi = 30°$ and again becomes quantized mode when $\phi$ is increased to $45°$ but the other anisotropic modes M2, M3 and M8 remain almost unaltered with the variation of $\phi$. The origin of the observed two-fold rotational anisotropy can be explained by considering the boundary effect coming from the rectangular shape of the boundary of the samples which has been confirmed from the angular dispersion spectra of NiFe thin film measured at $H_{ext}$ = 800 Oe (See Supplemental Material, Figure S5). To unravel the reason behind the presence of the six-fold anisotropic behaviour observed in all the lattices, the variation of the internal field ($B_{in}$) for D1-D4 is calculated at various orientations of the bias field by keeping its strength fixed to $H_{ext}$ = 800 Oe in the region marked by the blue dotted box as shown in M2 of D1 at $\phi$ = 15° in Fig. 6(a). It is evident from Fig. 6(c) that the variation of $B_{in}$ with $\phi$ indeed possesses a periodic six-fold rotational symmetry in case of all the samples, where this anisotropic contribution increases systematically from D1 to D4. Consequently, this is reflected as the anisotropic behaviour of the SW mode frequencies in the angular dispersion behaviour obtained for these lattices.

## IV. CONCLUSION

In summary, we have investigated the evolution of magnetization dynamics of hexagonally arranged NiFe circular antidot arrays with varying antidot diameter having same lattice spacing by controlling the bias magnetic field strength and its in-plane orientation using broadband ferromagnetic resonance technique. The field dispersion spectra of these systems reveal that the SW dynamics get drastically modified as the antidot diameter is varied. Rich multimodal SW spectra are obtained for the highest antidot diameter, whereas the number of SW modes reduces systematically with the decreasing antidot diameter. Moreover, a crossover between two higher frequency SW modes is observed for the lattices with higher hole diameter when the strength of the bias field is reduced. The simulated static magnetic configurations along with the power and phase profiles unravel the spatial distribution of the observed SW modes which confirms the formation of SW quantization laterally as well as diagonally inside the array having higher hole diameter at low bias field due to the strong overlapping of demagnetization regions between the antidots. As a result, the internal field is reduced significantly with the enhancement in antidot diameter. The variation of magnonic spectra with the in-plane orientation for all the samples shows the presence of two anisotropic SW modes both in opposite phase to each other, with six-fold rotational symmetry which is strongly modulated when the hole diameter is increased. Interestingly, for intermediate antidot diameters, the lowest frequency SW mode apparently shows a mode hopping-like behaviour with 60° periodicity but a closer look reveals a parametric splitting of mode. The phase maps of these samples unveil an interesting conversion from extended nature to quantized standing wave pattern or vice-versa in most of these anisotropic SW modes with the modulation of in-plane orientation. Further, the variation of internal field with the in-plane orientation confirms the presence of six-fold anisotropy which is strongly modulated when the diameter of the antidots is modified. Thus, the variation in both the antidot size and the magnetic field orientation demonstrate active methods essentially leading to a modulation in the profile of a periodically varying SW channel which may subsequently determine the SW frequency dispersion. This property can be implemented in dynamic spin-wave filters and magnonic waveguides in the gigahertz frequency range. Also, the observed mode hopping-like behaviour can be utilized in non-linear magnonic devices or coupled waveguides analogous to opto-electronic devices in photonics. Therefore, the observed tunability of the magnetization dynamics with the antidot size as well as the strength and orientation of the in-plane bias field plays a crucial role from a fundamental scientific viewpoint as well as in terms of the nanoscale magnonic crystal-based technology.

## V. ACKNOWLEDGMENTS

The authors gratefully acknowledge the financial supports from the Department of Science and Technology, Government of India under grant no. SR/NM/NS-09/2011 and S. N. Bose National Centre for Basic Sciences for grant no. SNB/AB/12-13/96 and SNB/AB/18-19/211. S. C. acknowledges S. N. Bose National Centre for Basic Sciences for financial assistance.

LIST OF FIGURES:

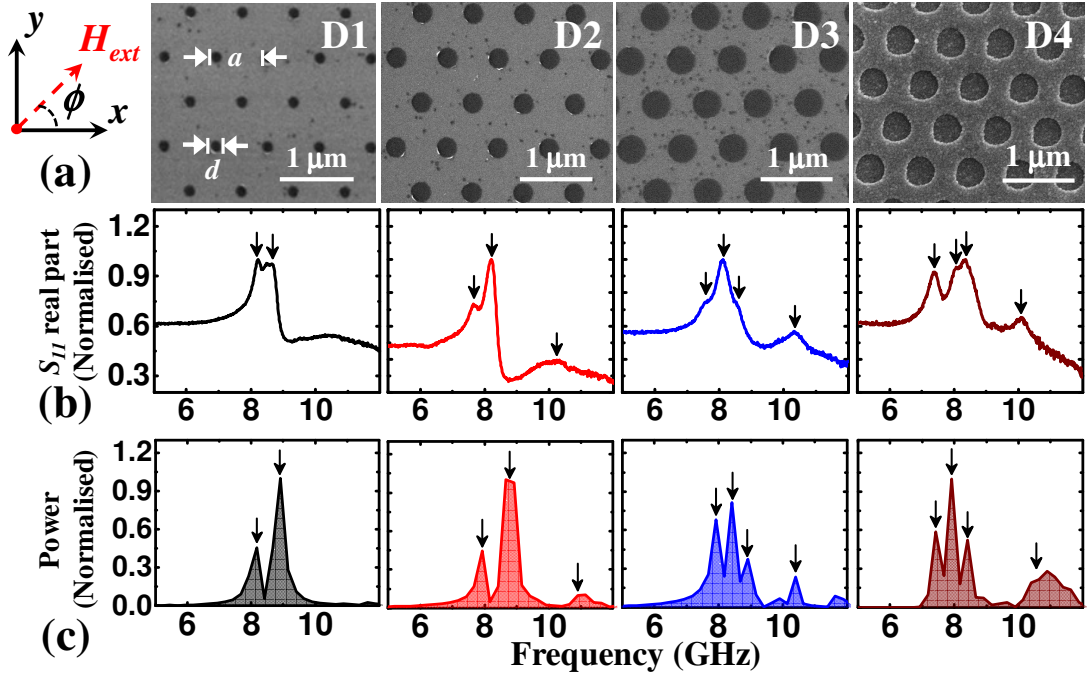

FIG. 1. (a) Scanning electron micrographs of circular-shaped $Ni_{80}Fe_{20}$ (NiFe) antidots (ADs) arranged in hexagonal lattices of constant lattice spacing $a$ = 700 nm with variable antidot (hole) diameter $d$ = 140 (D1), 240 (D2), 340 (D3) and 440 (D4) nm. (b) Real parts of the forward scattering parameter ($S_{11}$) representing the FMR spectra for all four samples at bias magnetic field $H_{ext}$ = 800 Oe applied at an azimuthal angle $\phi$ = 0° and the observed SW modes were marked by down arrows. (c) Corresponding simulated SW spectra of four different lattices at $H_{ext}$ = 800 Oe applied at $\phi$ = 0° and the arrows represent different SW modes. The orientation of the bias magnetic field $H_{ext}$ is shown at the top left of the figure.

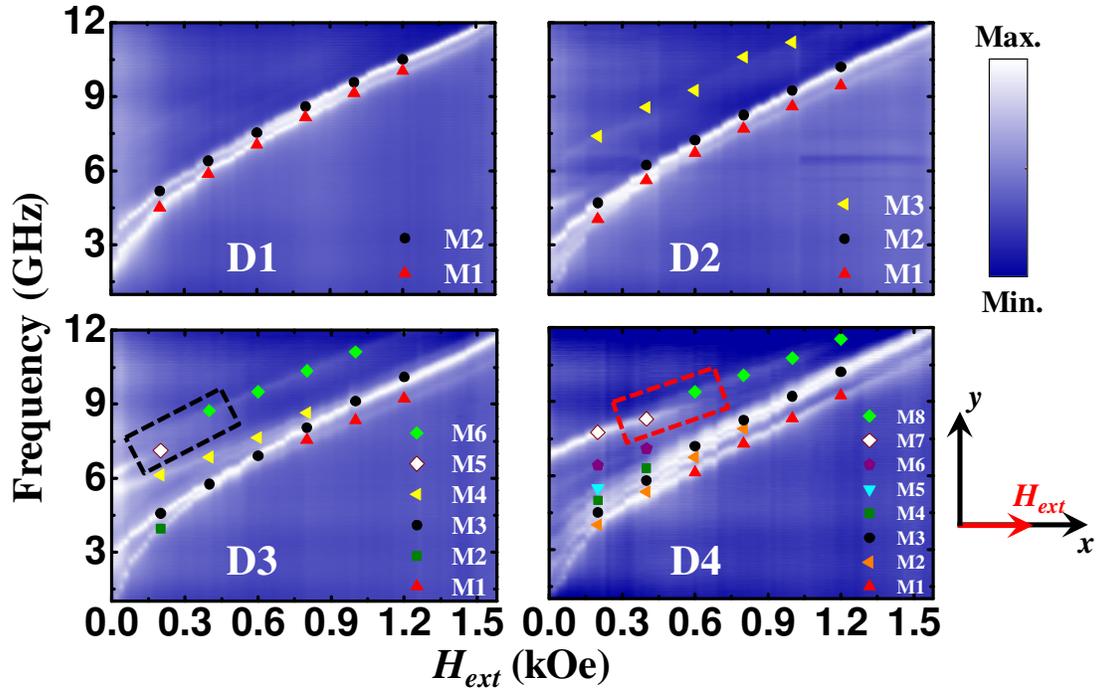

FIG. 2. Bias field ($H_{ext}$) dependent SW absorption spectra of NiFe ADLs with different antidot diameters (D1-D4) are shown at $\phi = 0°$. The surface plots correspond to the experimental results, while the symbols represent the simulated data. The black and red dotted boxes represent the crossover between two higher frequency branches, *i.e.* M5 and M6 in D3 and M7 and M8 in D4, respectively. The color map for the surface plots and the schematic of $H_{ext}$ are given at the right side of the figure.

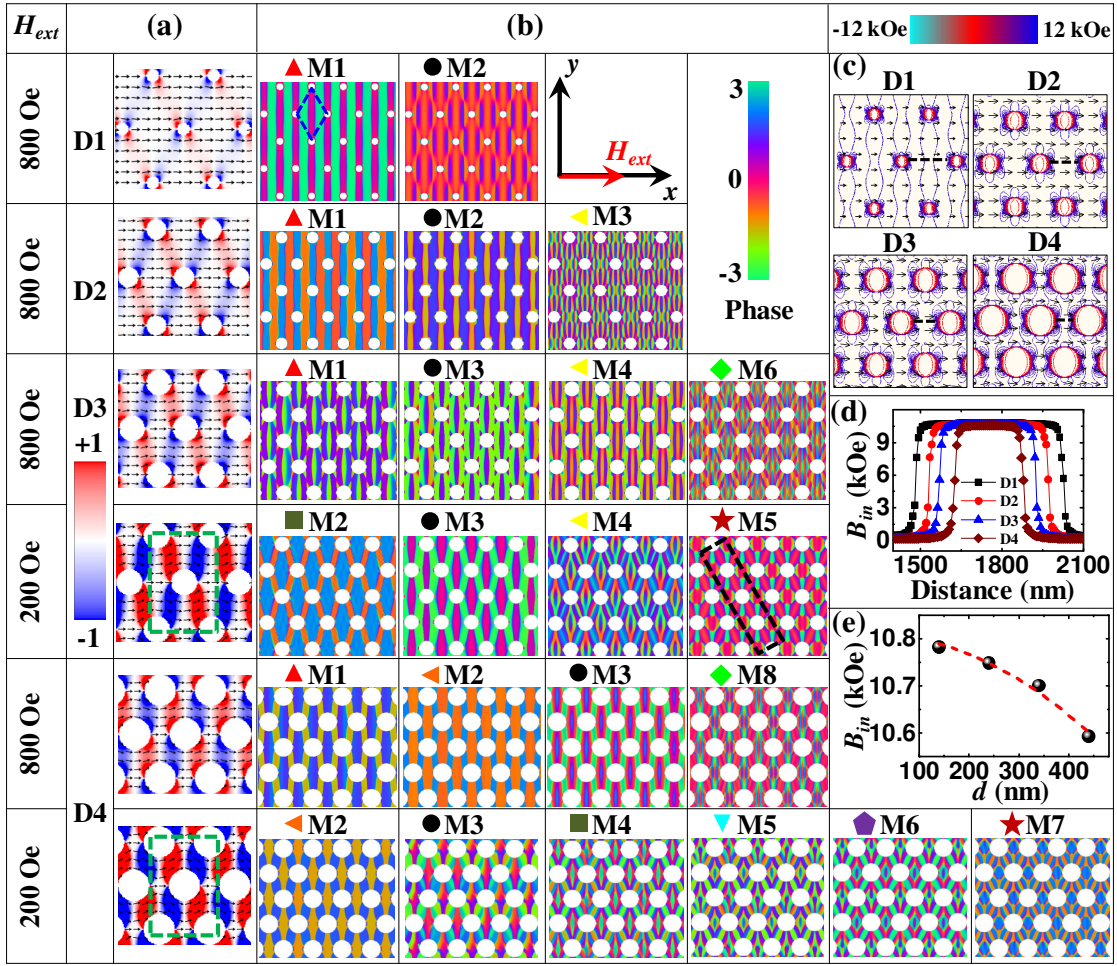

FIG. 3. (a) Simulated static magnetic configurations of NiFe ADLs are shown for $H_{ext}$ = 800 Oe in D1-D4 and $H_{ext}$ = 200 Oe in D3 and D4, respectively, at $\phi$ = 0°. The color map is given on the left side of the figure. (b) Simulated spatial distributions of the phase profiles corresponding to different SW modes obtained at $H_{ext}$ = 800 Oe in D1-D4 and $H_{ext}$ = 200 Oe in D3 and D4, respectively, at $\phi$ = 0°. The color map for the phase distributions and the schematic of $H_{ext}$ are shown on the right side of the figure. (c) Contour plots of the simulated magnetostatic field distributions in D1-D4 and the corresponding color map is given at the top right corner of the figure. (d) Linescans of the simulated internal field ($B_{in}$) taken between two consecutive antidots along the black dotted lines as shown in (c) for D1-D4. (e) The variation of $B_{in}$ with the antidot diameter $d$ (black circular symbols: micromagnetic symbols; red dotted line: fitted curve).

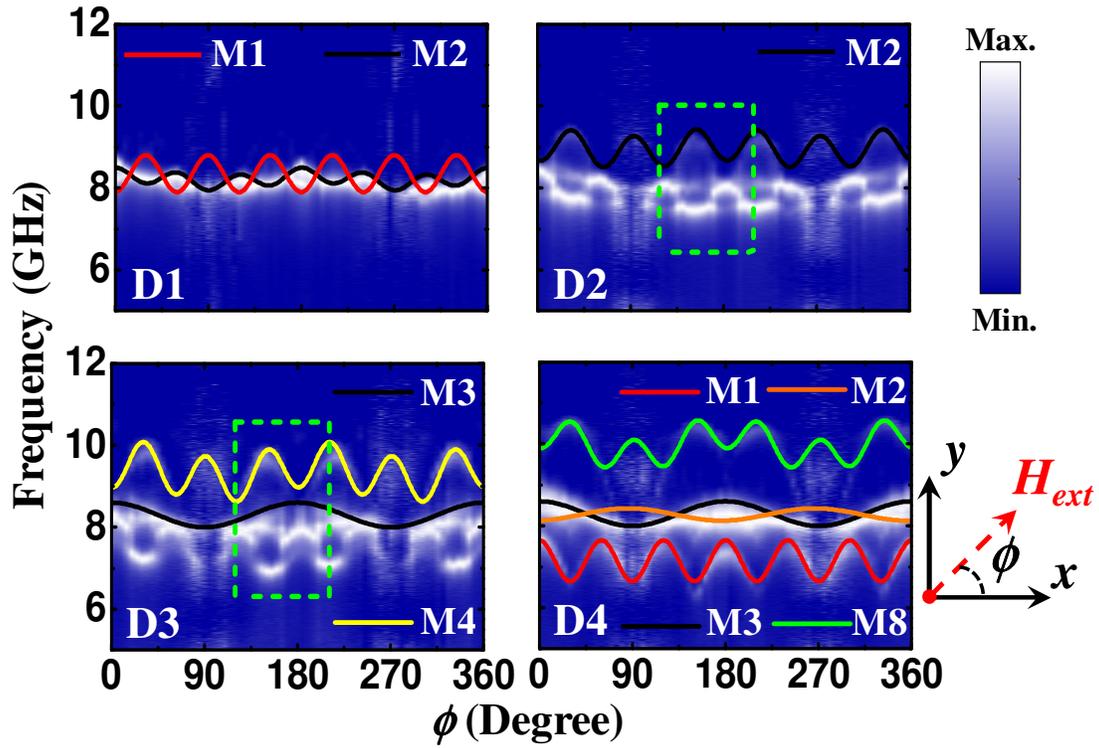

FIG. 4. Variation of SW frequency with the azimuthal angle ($\phi$) varying from 0° to 360° for NiFe ADLs with various antidot diameter (D1-D4) at $H_{ext}$ = 800 Oe. The surface plots represent the experimental results while the solid lines describe the sinusoidal fits to the observed anisotropic SW modes in all of the samples (D1-D4). The color map associated with the surface plots and the schematic of the orientation of $H_{ext}$ are shown on the right side of the figure.

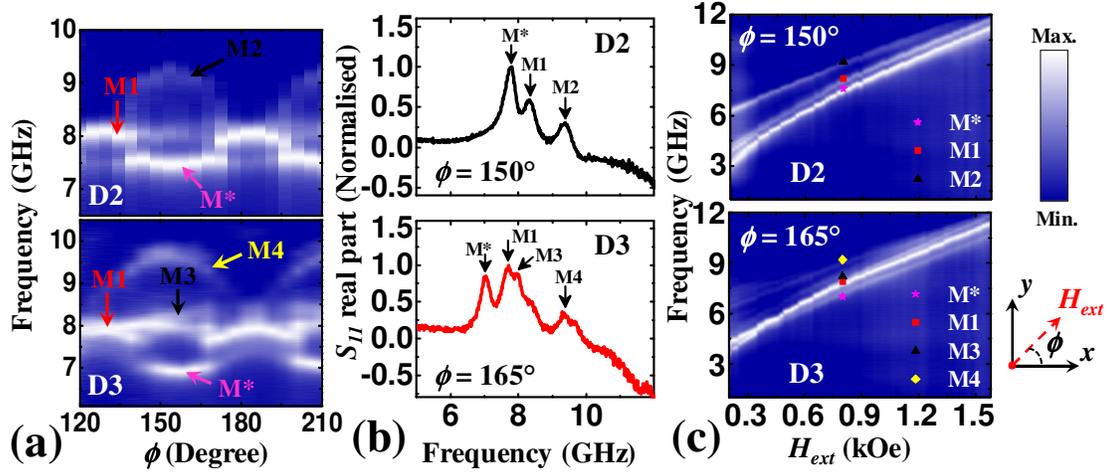

FIG. 5. (a) Angular dispersion of FMR spectra of D2 and D3 for $120° \leq \phi \leq 210°$ obtained from Fig. 4 as marked by green dotted boxes in D2 and D3. (b) Real parts of the forward scattering parameter ($S_{11}$) representing the FMR spectra for D2 and D3 at bias magnetic field $H_{ext}$ = 800 Oe applied at an azimuthal angle $\phi$ = 150° and 165°, respectively. The observed SW modes were marked by down arrows. (b) Bias field ($H_{ext}$) dependent SW absorption spectra of D2 at $\phi$ = 150° and D3 at $\phi$ = 165° are shown with $H_{ext}$ = 800 Oe. The surface plots correspond to the experimental results, while the symbols represent the simulated data. The color map for the surface plots and the schematic of $H_{ext}$ are given on the right side of the figure.

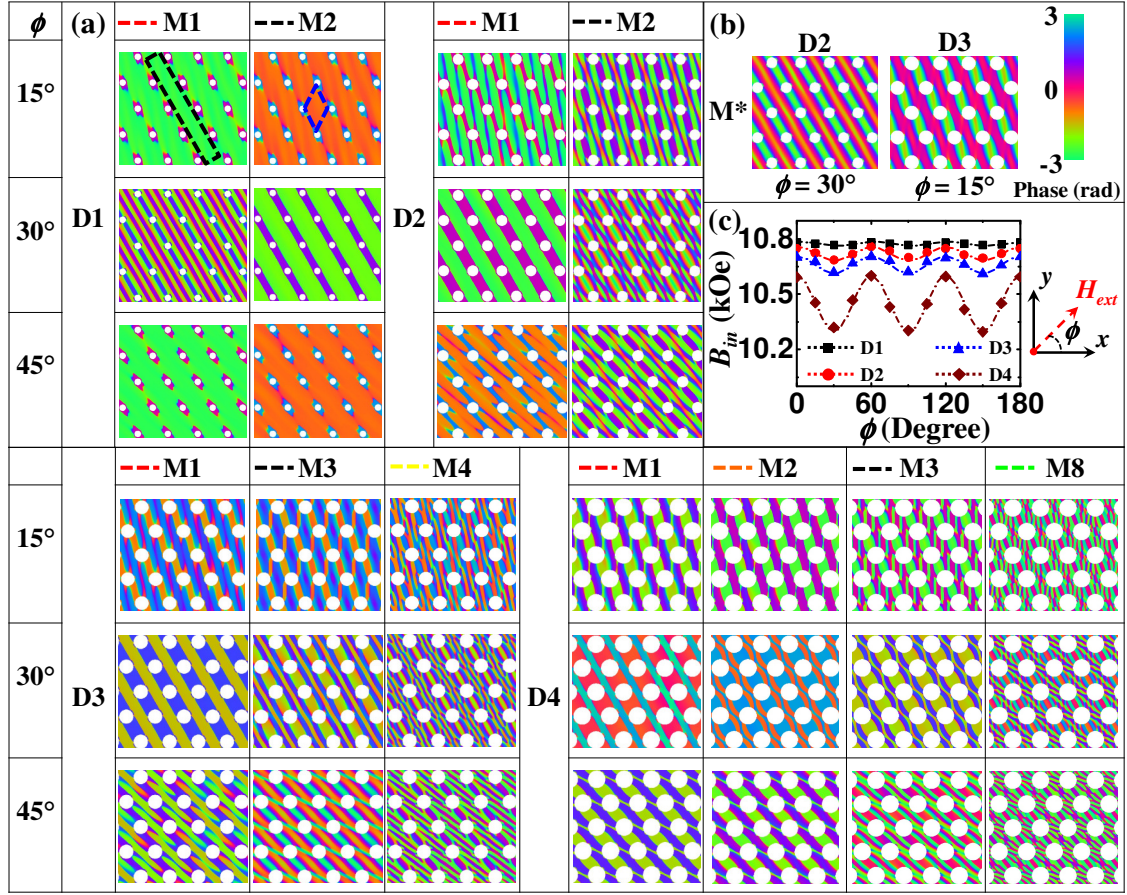

FIG. 6. (a) Simulated spatial distributions of the phase profile corresponding to different anisotropic SW modes obtained in D1-D4 with $H_{ext}$ = 800 Oe at $\phi$ = 0°, 15° and 30°, respectively. The black dotted box inside mode 1 (M1) of D1 at $\phi$ = 15° represents the nature of extended SW mode in a direction perpendicular to $H_{ext}$. (b) Simulated phase maps of the additional SW mode (M*) obtained in D2 at $\phi$ = 30° and D3 at $\phi$ = 15°, respectively with $H_{ext}$ = 800 Oe. The color map for the phase distributions and the schematic of $H_{ext}$ are shown on the top right corner of the figure. (c) Evolution of the simulated internal field ($B_{in}$) values in D1-D4 with varying $\phi$ at $H_{ext}$ = 800 Oe obtained by taking linescans in the region marked by blue dotted box inside mode 2 (M2) of D1 at $\phi$ = 15°.